%% file: root.tex
\documentclass{ifacconf}

\usepackage{graphicx}  
\usepackage{natbib}  
\input{customize}
\input{glossary.tex}
\begin{document}
\begin{frontmatter}

\title{Offline Uncertainty Sampling in Data-driven Stochastic MPC\thanksref{footnoteinfo}}

\thanks[footnoteinfo]{\copyright~2023 the authors. This work has been accepted to IFAC for publication under a Creative Commons Licence CC-BY-NC-ND}

\author[First]{Johannes Teutsch} 
\author[First]{Sebastian Kerz} 
\author[First]{Tim Br\"udigam}
\author[First]{Dirk Wollherr}
\author[First]{Marion Leibold}

\address[First]{Technical University of Munich, Department of Computer Engineering, Chair of Automatic Control Engineering (LSR), Theresienstra{\ss}e 90, 80333 Munich, Germany (e-mail: { \{johannes.teutsch, s.kerz, tim.bruedigam, dw, marion.leibold\}@tum.de})}

\begin{abstract}                
In this work, we exploit an offline-sampling based strategy for the constrained data-driven predictive control of an unknown linear system subject to random measurement noise. The strategy uses only past measured, potentially noisy data in a non-parametric system representation and does not require any prior model identification. The approximation of chance constraints using uncertainty sampling leads to efficient constraint tightening. Under mild assumptions, robust recursive feasibility and closed-loop constraint satisfaction is shown. In a simulation example, we provide evidence for the improved control performance of the proposed control scheme in comparison to a purely robust data-driven predictive control approach.
\end{abstract}


\end{frontmatter}

\input{1_intro}

\input{2_prelim}

\input{3_problem}

\input{4_method}

\input{5_eval}

\input{6_conclusion}

\def\bibfont{\small}
\bibliography{mybib}   

\appendix
\input{appendix}

\end{document}

%% file: customize.tex
\usepackage{amsmath,amssymb,amsfonts}
\usepackage{siunitx}
\usepackage{bm}
\usepackage{graphicx}
\usepackage{xcolor}
\usepackage{url}
\graphicspath{{pics/}}

\newcommand{\mat}[1]{\ensuremath{\begin{bmatrix} #1 \end{bmatrix}}}
\newcommand{\vc}[1]{\ensuremath{\begin{pmatrix} #1 \end{pmatrix}}}

\newcommand{\eps}{\varepsilon}
\renewcommand{\rho}{\varrho}
\renewcommand{\Pr}[1]{\textnormal{Pr}\left(#1\right)}
\newcommand{\E}[1]{\textnormal{E}\left(#1\right)}

\newcommand{\norm}[1]{\left\lVert #1 \right\rVert}
\newcommand{\norminf}[1]{\left\lVert #1 \right\rVert_{\infty}}

\newcommand{\diag}[1]{\textnormal{diag}\left( #1 \right)}
\renewcommand{\text}[1]{\textnormal{#1}}
\newcommand{\conv}[1]{\textnormal{conv}\left(#1\right)}

%% file: glossary.tex
\usepackage[acronym, style=alttree, toc=true, shortcuts, xindy, nomain, nonumberlist]{glossaries}

\newglossary{symbols}{sym}{sbl}{List of Symbols}
\newglossary{notation}{not}{nt}{Notation}

\glsaddkey{dimension}{\glsentrytext{\glslabel}}{\glsentryd}{\GLsentryd}{\glsd}{\Glsd}{\GLSd}

\newacronym{MPC}{MPC}{model predictive control}
\newacronym{iss}{ISS}{input-to-state stability}

\newglossaryentry{x}{type=symbols,
	sort={x},
	dimension={\ensuremath{ \mathbb{R}^{n} }},
	name={\ensuremath{\bm{x}}},
	description={State}
}

\newglossaryentry{xh}{type=symbols,
	sort={x},
	dimension={\ensuremath{ \mathbb{R}^{n} }},
	name={\ensuremath{\hat{\bm{x}}}},
	description={Measured state}
}

\newglossaryentry{u}{type=symbols,
	sort={u},
	dimension={\ensuremath{ \mathbb{R}^{m} }},
	name={\ensuremath{\bm{u}}},
	description={Input}
}

\newglossaryentry{y}{type=symbols,
	sort={y},
	dimension={\ensuremath{ \mathbb{R}^{n_y} }},
	name={\ensuremath{\bm{y}}},
	description={Output}
}

\newglossaryentry{d}{type=symbols,
	sort={d},
	dimension={\ensuremath{ \mathbb{D} }},
	name={\ensuremath{\bm{d}}},
	description={Disturbance}
}

\newglossaryentry{w}{type=symbols,
	sort={w},
	dimension={\ensuremath{ \mathbb{W} }},
	name={\ensuremath{\bm{w}}},
	description={General / extended disturbance}
}

\newglossaryentry{eps}{type=symbols,
	sort={e},
	dimension={\ensuremath{ \mathbb{E}_{\eps}}},
	name={\ensuremath{\bm{\eps}}},
	description={Measurement noise}
}

\newglossaryentry{epsbnd}{type=symbols,
	sort={e},
	dimension={\ensuremath{ \mathbb{E}_{\eps}}},
	name={\ensuremath{\overline{\eps}}},
	description={Measurement noise}
}

\newglossaryentry{ud}{type=symbols,
	sort={u},
	dimension={\ensuremath{ \mathbb{R}^{m} }},
	name={\ensuremath{\bm{u}^{\textnormal{d}}}},
	description={Input data}
}

\newglossaryentry{epsd}{type=symbols,
	sort={\eps},
	dimension={\ensuremath{ \mathbb{R}^{n} }},
	name={\ensuremath{\bm{\eps}^{\textnormal{d}}}},
	description={Measurement noise data}
}

\newglossaryentry{xd}{type=symbols,
	sort={x},
	dimension={\ensuremath{ \mathbb{R}^{n} }},
	name={\ensuremath{\bm{x}^{\textnormal{d}}}},
	description={State data}
}

\newglossaryentry{xhd}{type=symbols,
	sort={x},
	dimension={\ensuremath{ \mathbb{R}^{n} }},
	name={\ensuremath{\hat{\bm{x}}^{\textnormal{d}}}},
	description={State data}
}

%% file: 1_intro.tex
\section{INTRODUCTION}
Model Predictive Control (MPC) is an optimization based control method that relies on repeatedly solving an optimal control problem (OCP) over a predefined prediction horizon, allowing to take performance criteria and constraints on system variables into account~\citep{mayne2014model}. However, closed-loop performance and constraint satisfaction heavily depend on the accuracy of the prediction model. When models are uncertain and identification is costly, integrating data into MPC schemes promises to improve control performance.

An appealing alternative to classical system models is provided by a result from behavioral systems theory known as Willems' \textit{fundamental lemma}~\citep{willems2005note}, which has recently received increasing interest in research. The lemma states that Hankel matrices, built from an input-output trajectory with a persistently exciting (PE) input signal, span the space of all possible input-output trajectories of the underlying linear time-invariant (LTI) system. The resulting data-driven system representation allows for solving control and analysis problems directly from data, without requiring model identification~\citep{de2019formulas,datainformativity,berberichECC2020}, and in particular to substitute the prediction model within MPC for finding optimal future input-output sequences~\citep{coulson2019data,berberich2020data}. A concise and comprehensive recent review is provided by~\cite{behavioraltheory2021}.

Although it has been shown that data-driven MPC (DD-MPC) schemes can perform remarkably well even for nonlinear systems~\citep{deepcQuadcop,berberichAT}, the fundamental lemma itself only considers deterministic LTI systems and noise-free measurement data. In order to deal with inconsistent data in practice, regularized slack variables are introduced in the OCP~\citep{coulson2019data,berberich2020data}. In \cite{berberich2020data}, a DD-MPC scheme for systems subject to bounded measurement noise with guarantees on stability and robustness was presented, providing the first theoretical analysis of closed-loop properties that result from a purely DD-MPC scheme. Robust constraint tightening considering noisy data for the scheme from~\cite{berberich2020data} is presented in~\cite{berberich2020robust}; however, a restrictive terminal equality constraint is implemented and the constraint tightening is based on conservative estimates of parameter bounds. A data trajectory resulting from systems subject to stochastic uncertainties with unknown probability distribution was considered by \cite{coulson2021distributionally}. Strong probabilistic guarantees on out-of-sample performance and distributionally robustness are shown, albeit only the open-loop is considered and closed-loop properties like recursive feasibility are left for future work. So far, research on DD-MPC has either not considered noisy data or has treated measurement noise robustly, without incorporating potentially known probabilistic properties of the noise. 

Stochastic MPC (SMPC) has emerged as a framework to systematically incorporate probabilistic descriptions of uncertainties for effective constraint tightening~\citep{mesbah2016}. By introducing \textit{chance constraints}, i.e., requiring state or output constraints to be satisfied with a pre-specified probability level, SMPC allows for a systematic trade-off between control performance and constraint satisfaction. This is particularly important for MPC of uncertain systems when optimal performance requires operation near constraint boundaries in applications where rare or transient constraint violations are acceptable, such as in  electric grids~\citep{JiangEtalDong2019} or finance \citep{GrafPlessenEtalBemporad2019}. For safety-critical applications, safety guarantees are enabled by employing failsafe or robust backup plans \citep{WabersichEtalZeilinger2021, BruedigamEtalLeibold2021b, BruedigamEtalLeibold2021c}. 

A major challenge in SMPC is to reformulate the chance constraint into a deterministic expression for tractability of the OCP.
Sampling-based approaches provide an appealing remedy, as they are easy to implement, independent of the underlying probability distribution, and also allow for nonlinearity of the uncertainties in the system dynamics~\citep{lorenzen2017stochastic}. Approaches that exploit online-sampling of the uncertainties for the scenario approximation of SMPC problems are known as Scenario MPC~\citep{schildbach2014scenario}. While online-sampling comes with reduced sample complexity, offline-sampling allows for a reduced online computational load by removing redundant constraints offline, as well as for a guarantee of recursive feasibility by introducing a constraint on the first predicted step, as proposed by~\cite{lorenzen2017stochastic} for known system matrices subject to parametric uncertainties. On the other hand, sampling-based approaches can only guarantee chance constraint satisfaction with confidence, and the computational load increases drastically with the dimension of the system~\citep{schildbach2014scenario}.

In the data-driven setting, a reformulation of the chance constraint may be achieved by leveraging polynomial chaos expansion \citep{pan2021stochastic} or employing stochastic tubes~\citep{kerz2023datadriven}. However, both works consider systems subject to additive stochastic disturbances, and assume the available (measured) data to be exact.

\textit{Contribution:}
In this work, we present a novel strategy for the data-driven stochastic predictive control of unknown LTI systems subject to measurement noise. 
Based on Willems' fundamental lemma, the algorithm requires only a single initially measured PE input-state trajectory, and leverages chance constraints to efficiently control against both noise in the data and in online state measurements.

In order to render the chance constraints tractable, we employ an offline-sampling based approach similar to~\cite{lorenzen2017stochastic}.
Similar to \cite{lorenzen2017stochastic} and \cite{kerz2023datadriven}, a constraint on the first predicted step guarantees recursive feasibility and closed-loop constraint satisfaction without the need of restrictive terminal equality constraints and slack variables. 
In contrast to~\cite{lorenzen2017stochastic}, the proposed approach further considers online measurement noise and does not require any model knowledge.
To the best of our knowledge, this it the first work that integrates sampling-based constraint tightening into a DD-MPC scheme in order to safely cope with the uncertainty in the data-driven system representation, stemming from noisy system data.

\textit{Structure:}
This work is structured as follows. Section~\ref{sec:prelim} introduces preliminary results related to the data-driven framework and SMPC, before the problem is formulated
in Section~\ref{sec:problem}. The proposed sampling-based data-driven SMPC (DD-SMPC) algorithm is presented in Section~\ref{sec:method}. In Section~\ref{sec:eval}, we provide a simulation example to evaluate our approach. The work closes with a conclusion in Section~\ref{sec:conclusion}. 

\textit{Notation:}
We write $\bm{0}$ for any zero matrix or vector, and we abbreviate the set of integers $\left\{a,\,\ldots,\, b\right\}$ by $\mathbb{N}_a^b$. Sequences of vectors are shortened to $\bm{s}_{[1,\,k]} = \left\{\bm{s}_1,\,\ldots,\,\bm{s}_k\right\}$, while the vector $\underline{\bm{s}}_{[1,\,k]} = \left(\bm{s}_1^{\top},\,\ldots,\, \bm{s}_k^{\top}\right)^{\top}$ denotes the result from stacking the elements of $\bm{s}_{[1,\,k]}$. For a given state $\bm{x}_k$, we write $\bm{x}_{l|k}$ for the predicted state $l$ steps ahead. The Moore-Penrose pseudo-rightinverse of a matrix $\bm{D}$ is defined as $\bm{D}^{\dagger} = \bm{D}^{\top}\left(\bm{D}\bm{D}^{\top}\right)^{-1}$. The matrix $\left[\bm{D}\right]_{[a,\,b]}$ consists of all rows starting from the $a$-th row to the $b$-th row of $\bm{D}$. The probability measure is defined by $\Pr{\cdot}$, whereas the expectation operator is denoted as $\E{\cdot}$. For any sequence of vectors $\bm{s}_{[1,\,N]}$ with length $N$, the corresponding Hankel matrix $\bm{H}_{L}\left(\bm{s}_{[1,\,N]}\right)$ of order $L \le N$ is defined as
\begin{equation} \label{def:hankel}
\bm{H}_{L}\left(\bm{s}_{[1,\,N]}\right) = \mat{\underline{\bm{s}}_{[1,\,L]},\,\underline{\bm{s}}_{[2,\,L+1]},\,\ldots,\,\underline{\bm{s}}_{[N-L+1,\,N]}}.
\end{equation}

%% file: 2_prelim.tex
\section{PRELIMINARIES} \label{sec:prelim}
In this section, we provide preliminary results of data-driven control based on behavioral systems theory and sampling-based SMPC.
\subsection{Data-driven system representation}
Consider the standard definition of persistency of excitation as given in the following \citep{willems2005note}.
\begin{defn}[Persistency of excitation] \label{def:persistency}
	~~\,A sequence $\bm{s}_{[1,\,N]}$ of length $N$ with $\bm{s}_i \in \mathbb{R}^m$ is PE of order $L$ if the Hankel matrix $\bm{H}_{L}\left(\bm{s}_{[1,\,N]}\right)$ has full rank~$mL$.
\end{defn}
The following result is known as Willems' \textit{fundamental lemma} \citep{willems2005note}, which is a result from behavioral systems theory and lays the foundation for describing system behavior without model-knowledge in this work. We state the lemma in its state-space form.
\begin{lem}[Willems' lemma~\citep{de2019formulas}]\label{lem:fundamental}~\linebreak
	Consider a controllable LTI system of the form $\gls{x}_{k+1} = \bm{A} \gls{x}_{k} + \bm{B} \gls{u}_{k}$ with given data sequences $\gls{ud}_{[1,\,N]}$, $\gls{xd}_{[1,\,N]}$. If $\gls{ud}_{[1,\,N]}$ is PE of order $n+L+1$, then any input-state sequence $\left(\gls{u}_{[k,\,k+L]},\gls{x}_{[k,\,k+L]}\right)$ is a valid trajectory of the system if and only if there exists an $\bm{\alpha} \in \mathbb{R}^{N-L}$ such that
	\begin{equation} \label{eq:fundlemm_eq}
		\vc{\underline{\gls{u}}_{[k,\,k+L]}\\ \underline{\gls{x}}_{[k,\,k+L]}} = \mat{\bm{H}_{L+1}\left(\gls{ud}_{[1,\,N]}\right) \\ \bm{H}_{L+1}\left(\gls{xd}_{[1,\,N]}\right)} \bm{\alpha}.
	\end{equation}
\end{lem}
Lemma~\ref{lem:fundamental} describes an implicit and non-parametric system representation that is purely based on data. 
\begin{rem}
Lemma~\ref{lem:fundamental} is only valid if noise-free state data~$\gls{xd}_{[1,\,N]}$ are used in \eqref{eq:fundlemm_eq}. In our setting, we assume to only have access to noisy state data $\gls{xhd}_{[1,\,N]} = \gls{xd}_{[1,\,N]} + \gls{epsd}_{[1,\,N]}$ with the noise $\gls{epsd}_{[1,\,N]}$, which introduces uncertainty for trajectories computed from~\eqref{eq:fundlemm_eq} as the noise is multiplied by the parameter~$\bm{\alpha}$. More details are given in Section~\ref{sec:problem}.
\end{rem}
\subsection{Sampling-based SMPC} \label{sec:prelim_sampling}
In SMPC, hard state constraints $\gls{x}_{k} \in \mathbb{X}$ are replaced by chance constraints of the form~\citep{mesbah2016}
\begin{equation} \label{eq:statecons_prob}
    \Pr{\gls{x}_{k} \in \mathbb{X}} \ge p ~~ \forall k \in \mathbb{N},
\end{equation}
with the risk parameter $p \in (0,1)$. Chance constraints lead to less conservative control actions and allow for a deliberate trade-off between performance and constraint satisfaction with the risk parameter $p$. The control becomes more conservative as $p$ approaches $1$; a choice of $p=1$ is equivalent to robust constraint satisfaction.

For tractability of the resulting OCP in the predictive control scheme, the chance constraint \eqref{eq:statecons_prob} needs to be reformulated into a deterministic expression. In case of nonlinear influence of the uncertainties in the system dynamics or non-Gaussian distributions, sampling-based approaches can be used to deterministically approximate the chance constraint~\eqref{eq:statecons_prob}. In the following, we will present relevant theory for offline constraint sampling.

Let $\bm{\theta}\in\mathbb{R}^{r \times d}$ and $\bm{\phi} \in \mathbb{R}^{r}$ be multivariate random variables, and let the set of decision variables $\bm{\xi}$ for which the constraint ${\bm{\theta} \bm{\xi} \le \bm{\phi}}$ holds with at least probability $p$ be
\begin{equation}
    \mathbb{S}^P = \left\{ \bm{\xi} \in \mathbb{R}^d ~\left|~ \Pr{\bm{\theta} \bm{\xi} \le \bm{\phi}} \ge p \right. \right\}.
\end{equation}
Furthermore, for $i \in \mathbb{N}_1^{N_s}$, let $\bm{\theta}^{(i)}$ and $\bm{\phi}^{(i)}$ be $N_s$ independent and identically distributed (iid) samples of $\bm{\theta}$ and $\bm{\phi}$. We can then define the sampled set as
\begin{equation}
    \mathbb{S}^S = \left\{ \bm{\xi} \in \mathbb{R}^d ~\left|~ {\bm{\theta}^{(i)}} \bm{\xi} \le \phi^{(i)}, ~~ i \in \mathbb{N}_1^{N_s}\right. \right\}.
\end{equation}
The following proposition is a result from statistical learning theory~\citep{alamo2009randomized}, which lets us determine the required number of samples $N_s$ (i.e., the sample complexity) for which the sampled set $\mathbb{S}^S$ is a subset of the chance constraint set $\mathbb{S}^P$ with a predefined confidence $\beta$. 
\begin{prop}[Sampling subset~\citep{lorenzen2017stochastic}] \label{prop:sampling}
    For any risk parameter $p \in (0,1)$, confidence level $\beta \in (0,1)$,~and sample complexity $N_s$ that satisfies
    \begin{equation} \label{eq:samplecomplexity}
        N_s \ge \tilde{N}(d,p,\beta) = \frac{5}{1-p}\left( \ln \frac{4}{1-\beta} + d \ln \frac{40}{1-p}\right),
    \end{equation}
    it holds that $\Pr{\mathbb{S}^S \subseteq \mathbb{S}^P} \ge \beta$, with $d = \dim{\bm{\xi}}$ for constraints of the form ${\bm{\theta}\bm{\xi}\le\bm{\phi}}$ that contain the origin.
\end{prop}
We can now elaborate on the problem setup that is addressed in this work and state all required assumptions.

%% file: 3_problem.tex
\section{PROBLEM SETUP} \label{sec:problem}
We consider a controllable discrete-time LTI system with unknown system matrices $\bm{A}$ and $\bm{B}$, i.e.,
\begin{subequations}\label{eq:system}
	\begin{align}
	\gls{x}_{k+1} &= \bm{A} \gls{x}_{k} + \bm{B} \gls{u}_{k}, \label{eq:system1}\\
	\gls{y}_{k} &= \gls{xh}_{k} := \gls{x}_{k} + \gls{eps}_{k},\label{eq:system2}
	\end{align}
\end{subequations}
where $\gls{x}_{k}\in \glsd{x}$ denotes the state and $\gls{u}_{k}\in \glsd{u}$ denotes the input. We consider the measured (noisy) state $\gls{xh}_{k}$ as the output $\gls{y}_{k}$ of the system, subject to the noise~$\gls{eps}_{k}$.
\begin{assum}[Measurement noise]\label{ass:noise}
    The noise $\gls{eps}_{k}$ is assumed to be the realization of a random variable that is distributed according to a known probability distribution function $\bm{f}_{\eps}\left(\cdot\right)$ and supported by the compact polytopic set
    \begin{equation} \label{eq:noiseset}
    	\glsd{eps} = \left\{\bm{\eps} \in \mathbb{R}^n ~\left|~ \bm{G}_{\eps} \bm{\eps} \le \bm{g}_{\eps} \right.\right\} \subset \mathbb{R}^n.
    \end{equation}
\end{assum}
Although the system matrices $\bm{A}$ and $\bm{B}$ are unknown, we assume to know an outer-bounding set $\mathbb{A}$ for the system matrices to later guarantee robust recursive feasibility. 
\begin{assum}[Outer-bound to system matrices]\label{ass:bound}
    A polytopic set
    $\mathbb{A} = \conv{\left\{\bm{A}_j, \bm{B}_j\right\}_{j \in \mathbb{N}_1^{N_c}}}$
    with $\left\{\bm{A},\bm{B}\right\} \in \mathbb{A}$ is known, where $N_c$ is the number of vertices of $\mathbb{A}$, and $\conv{\cdot}$ denotes the convex hull over the set of vertices.
\end{assum}
A possible strategy to derive such a set for the presented problem setup is given in Appendix~\ref{app:bounds}. Nevertheless, Assumption~\ref{ass:bound} can be relaxed to a confidence region $\mathbb{A}_{\tilde p}$ with ${\Pr{\left\{\bm{A},\bm{B}\right\}~\in~\mathbb{A}_{\tilde p}}~\ge~\tilde p}$, leading to probabilistic recursive feasibility guarantees~\citep{lorenzen2017stochastic}. Note that $\mathbb{A}$ is used to establish constraints only on the first predicted step, which allows for less conservative constraint tightening compared to standard robust MPC, e.g., \cite{mayne2014model}.

System~\eqref{eq:system1} is subject to polytopic state and input constraints for all time instants $k \in \mathbb{N}$, defined as
\begin{subequations}\label{eq:constraints}
	\begin{align}
	&\Pr{\gls{x}_{k} \in \mathbb{X}} \ge p, & \mathbb{X} = \left\{\bm{x} \in \mathbb{R}^n \hspace{2.5pt} ~\left|~ \bm{G}_x \bm{x} \le \bm{g}_x \right.\right\}, \label{eq:statecons} \\
	&\gls{u}_{k} \in  \mathbb{U}, & \mathbb{U} = \left\{\bm{u} \in \mathbb{R}^m ~\left|~ \bm{G}_u \bm{u} \le \bm{g}_u \right.\right\}, \label{eq:inputcons}
	\end{align}
\end{subequations}
where $\mathbb{X}$ and $\mathbb{U}$ are compact and contain the origin.

In our problem setting, we do not have access to the system matrices $\bm{A}$ and $\bm{B}$. Instead, we assume to have access to a persistently exciting data trajectory.
\begin{assum}[Persistently exciting data trajectory] \label{ass:data}
We assume to have access to one input-state trajectory $(\gls{ud}_{[1,N]}, \gls{xhd}_{[1,N]})$, where $\gls{ud}_{[1,\,N]}$ is assumed to be PE of order~$n+L+1$, and where the state data $\gls{xhd}_i = \gls{xd}_i + \gls{epsd}_i$ are affected by measurement noise according to Assumption~\ref{ass:noise}.
\end{assum}
Assumption~\ref{ass:data} is not restrictive in practice, since appropriate inputs $\gls{ud}_i$ can be chosen for the data collection. Note that we assume to only have access to noisy state data $\gls{xhd}_i$, i.e., the true state data $\gls{xd}_i$ and the noise realizations $\gls{epsd}_i$ are unknown.

We want to design a predictive control scheme for~\eqref{eq:system1}, but do not have access to a model to predict the future system behavior. Instead, we will directly make use of the given system data to predict trajectories using Lemma~\ref{lem:fundamental}.
However, these data are subject to measurement noise that introduces uncertainty into the prediction via~\eqref{eq:fundlemm_eq} as it is multiplied by the parameter~$\bm{\alpha}$. Currently available DD-MPC schemes based on Lemma~\ref{lem:fundamental} are able to ensure robustness \citep{berberich2020data} or distributional robustness \citep{coulson2021distributionally} with respect to noisy data, which can lead to overly conservative control performance. Using probabilistic knowledge about the noise~$\gls{eps}_{k}$ in system~\eqref{eq:system}, we will be able to exploit less conservative constraint tightening in this data-driven setting by employing ideas from sampling-based SMPC~\citep{lorenzen2017stochastic}.

\begin{prob}
The aim of this paper is the design of a predictive control scheme for system~\eqref{eq:system} that minimizes a quadratic cost function and guarantees hard constraint satisfaction for the input constraints~\eqref{eq:inputcons} and satisfaction of state constraints~\eqref{eq:statecons} up to a probability level specified by a risk parameter, cf.~\eqref{eq:statecons_prob}. For that, probabilistic knowledge about the measurement noise (Assumption~\ref{ass:noise}) is exploited. The system matrices in~\eqref{eq:system1} are unknown, but an outer-bounding set is given (Assumption~\ref{ass:bound}), as well as a persistently exciting data trajectory (Assumption~\ref{ass:data}).
\end{prob}

In the following section, Lemma~\ref{lem:fundamental} and Proposition~\ref{prop:sampling} are used to derive a DD-SMPC scheme based on offline uncertainty sampling to guarantee chance constraint satisfaction. Drawing the necessary number of samples offline allows to remove redundant constraints and guarantee recursive feasibility by adding a robust first-step constraint.

%% file: 4_method.tex
\section{SAMPLING-BASED DATA-DRIVEN SMPC} \label{sec:method}
In this section, we will first use the data-driven system representation to derive an explicit expression of the predicted state trajectory depending on uncertainty samples, such that application of the sampling-based approach from Section~\ref{sec:prelim_sampling} is possible. Then, we describe the constraint sampling procedure and the construction of the cost function, followed by the proposed control algorithm. We conclude this section with proofs of recursive feasibility and closed-loop constraint satisfaction.
\subsection{Data-driven Description of Sampled Trajectory}
For brevity, let us define the Hankel matrices
\begin{subequations}\label{eq:hankelmatrices}
    \begin{align}
        \bm{H}_u &= \bm{H}_{L+1}\left(\gls{ud}_{[1,\,N]}\right), &\bm{H}_{\hat{x}} &= \bm{H}_{L+1}\left(\gls{xhd}_{[1,\,N]}\right),\\
        \bm{H}_{x} &= \bm{H}_{L+1}\left(\gls{xd}_{[1,\,N]}\right), & \bm{H}_{\eps} &= \bm{H}_{L+1}\left(\gls{epsd}_{[1,\,N]}\right). \label{eq:noisehankel}
    \end{align}
\end{subequations}
Considering $\bm{H}_{\hat{x}} = \bm{H}_{x} + \bm{H}_{\eps}$, we can express the data-driven system representation~\eqref{eq:fundlemm_eq} in terms of the noisy state data and the unknown noise realization as
\begin{equation} \label{eq:sysrep1}
	\vc{\underline{\gls{u}}_{[k,\,k+L]} \\ \underline{\gls{x}}_{[k,\,k+L]}} = \mat{\bm{H}_{u}\\ \bm{H}_{\hat{x}} - \bm{H}_{\eps}} \bm{\alpha},
\end{equation}
where the state $\gls{x}_{k}$ at time instant $k$ can be expressed by using the noisy measured state and the unknown measurement noise, i.e., $\gls{x}_{k} = \gls{xh}_{k} - \gls{eps}_{k}$. Since the noise realization $\bm{H}_{\eps}$ that corrupted recorded data is also unknown, and robust approaches often lead to overly conservative control actions, we address the impact of the noise probabilistically using a sampling-based approach as described in Section~\ref{sec:prelim_sampling}. Particularly, given $N_s$ independently drawn uncertainty samples
$\bm{H}_{\eps}^{(i)}=\bm{H}_{L+1}\left(\gls{eps}^{\text{d},(i)}_{[1,\,N]}\right)$, $\gls{eps}_{k}^{(i)}$, ${i \in \mathbb{N}_1^{N_s}}$, a current state measurement $\gls{xh}_{k}$ and a future input sequence ${\bm{U}_{k} := \underline{\gls{u}}_{[0|k,\,L|k]}}$, we want all resulting state trajectory samples ${\bm{X}^{(i)}_{k} := \underline{\gls{x}}^{(i)}_{[0|k,\,L|k]}}$ computed via \eqref{eq:sysrep1} to satisfy \emph{hard constraints} at all prediction steps. This leads then to \emph{chance constraint} satisfaction with confidence according to Propositon~\ref{prop:sampling}. In the following, we provide an explicit expression for the sampled state trajectories $\bm{X}^{(i)}_{k}$ depending on $\gls{xh}_{k}$ and $\bm{U}_{k}$,
\begin{equation} \label{eq:samplepred}
    \bm{X}^{(i)}_{k} = \bm{M}^{(i)} \vc{\gls{xh}_{k} - \gls{eps}_{k}^{(i)}\\ \bm{U}_{k}},
\end{equation}
with a prediction matrix $\bm{M}^{(i)}$ that we will derive next.

For each sampled state trajectory $\bm{X}^{(i)}_{k}$, there exists $\bm{\alpha}^{(i)}$ such that~\eqref{eq:sysrep1} is satisfied, i.e.,
\begin{equation} \label{eq:sysrep1_sample}
	\vc{\bm{U}_{k}\\ \bm{X}^{(i)}_{k}} = \mat{\bm{H}_{u}\\ \bm{H}_{\hat{x}} - \bm{H}^{(i)}_{\eps}} \bm{\alpha}^{(i)},
\end{equation}
with $\gls{x}_{0|k}^{(i)} = \gls{xh}_{k} - \gls{eps}_{k}^{(i)}$. Using the given data $\bm{H}_{\hat{x}}$, $\bm{H}_u$ and the uncertainty samples $\bm{H}_{\eps}^{(i)}$, $\gls{eps}_{k}^{(i)}$, we can choose $\bm{\alpha}^{(i)}$ as
\begin{equation} \label{eq:alpha}
    \bm{\alpha}^{(i)} = \mat{ \left[\bm{H}_{\hat{x}} - \bm{H}_{\eps}^{(i)}\right]_{[1,\,n]} \\ \bm{H}_u}^{\dagger} \vc{\gls{xh}_{k} - \gls{eps}_{k}^{(i)} \\ \bm{U}_{k}},
\end{equation}
depending on the measured state $\gls{xh}_{k}$ and input sequence $\bm{U}_k$. Inserting~\eqref{eq:alpha} into~\eqref{eq:sysrep1_sample} delivers~\eqref{eq:samplepred}, with the matrix
\begin{equation} 
\vspace*{2mm}
    \bm{M}\left(\bm{H}_{\eps}^{(i)}\right) := \left(\bm{H}_{\hat{x}} - \bm{H}_{\eps}^{(i)}\right) \mat{ \left[\bm{H}_{\hat{x}} - \bm{H}_{\eps}^{(i)}\right]_{[1,\,n]} \\ \bm{H}_u}^{\dagger},
\end{equation}
$\bm{M}^{(i)} = \bm{M}\left(\bm{H}_{\eps}^{(i)}\right)$. From~\eqref{eq:samplepred}, we can determine the state trajectory for any sampled uncertainty realization. As a consequence, we can construct a cost function and reformulate the constraints~\eqref{eq:constraints} such that they only depend on the measured state $\gls{xh}_k$ and the input sequence $\bm{U}_{k}$, allowing for a minimal number of optimization variables.

\subsection{Cost Function and Offline Constraint Sampling} \label{sec:sampling}
Let us define the finite horizon cost function for the predictive control scheme as
\begin{align}\label{eq:cost_expected}
    &J_L\left(\gls{xh}_{k},\,\bm{U}_{k}\right)  = \notag \\[2mm] &\E{\sum\limits_{l=0}^{L-1} \left( \gls{x}_{l|k}^{\top} \bm{Q} \gls{x}_{l|k} + \gls{u}_{l|k}^{\top}\bm{R}\gls{u}_{l|k} \right) + \gls{x}_{L|k}^{\top}\bm{P}\gls{x}_{L|k}},
\end{align}
with $\gls{x}_{0|k} = \gls{xh}_{k} - \gls{eps}_{k}$, and where $\bm{Q}$, $\bm{P} \in \mathbb{R}^{n\times n}$ and ${\bm{R} \in \mathbb{R}^{m \times m}}$ are positive definite weighting matrices. From~\eqref{eq:samplepred}, we can retrieve the probabilistic state prediction ${\bm{X}_{k} := \underline{\gls{x}}_{[0|k,\,L|k]}}$  by substituting the samples $\bm{H}_{\eps}^{(i)}$, $\gls{eps}_{k}^{(i)}$ with corresponding random variables $\bm{H}_{\tilde\eps}$, $\tilde{\gls{eps}}$, i.e.,
\begin{equation} \label{eq:probpred}
    \bm{X}_{k} = \bm{M}\left(\bm{H}_{\tilde\eps}\right) \vc{\gls{xh}_{k} - \tilde{\gls{eps}}\\ \bm{U}_{k}}.
\end{equation}
We can then explicitly evaluate the expected cost~\eqref{eq:cost_expected} using~\eqref{eq:probpred}, resulting in the quadratic cost function
\begin{equation} \label{eq:cost_explicit}
    J_L\left(\gls{xh}_{k},\,\bm{U}_{k}\right)  = \vc{\gls{xh}_{k} \\ \bm{U}_{k}}^{\top} \bm{S} \vc{\gls{xh}_{k} \\ \bm{U}_{k}} + \bm{\gamma}^{\top}\vc{\gls{xh}_{k} \\ \bm{U}_{k}} + c,
\end{equation}
where $\bm{S},\,\bm{\gamma},\,c$ are defined as 
\begin{subequations} \label{eq:costparams}
    \begin{align}
        \bm{S} &= \E{\bm{M}\left(\bm{H}_{\tilde\eps}\right)^{\top} \tilde{\bm{Q}} \bm{M}\left(\bm{H}_{\tilde\eps}\right)} + \tilde{\bm{R}}, \label{eq:weight-S}\\
        \bm{\gamma} &= -\E{2 \bm{M}\left(\bm{H}_{\tilde\eps}\right)^{\top} \tilde{\bm{Q}} \bm{M}\left(\bm{H}_{\tilde\eps}\right) \vc{\tilde{\bm{\eps}}^{\top},\, \bm{0}}^{\top}},\\
        \bm{c} &= \E{\vc{\tilde{\bm{\eps}}^{\top},\, \bm{0}} \bm{M}\left(\bm{H}_{\tilde\eps}\right)^{\top}  \tilde{\bm{Q}} \bm{M}\left(\bm{H}_{\tilde\eps}\right) \vc{\tilde{\bm{\eps}}^{\top},\, \bm{0}}^{\top}},
    \end{align}
\end{subequations}
with the extended cost matrices $\tilde{\bm{Q}} = \diag{\bm{Q}, \dots, \bm{Q},\bm{P}}$, $\tilde{\bm{R}} = \diag{\bm{0}, \bm{R}, \dots, \bm{R},\bm{0}}$ of appropriate dimensions.
The expected value in \eqref{eq:costparams} can be evaluated by numerical integration to the desired accuracy, or by exploiting sample average approximation~\citep{kim2015guide}.
\begin{rem}
The parameter $c$ is a constant, thus, $c$ can be neglected in the optimization. Furthermore, the input $\gls{u}_{L|k}$ is not relevant for the optimization, but it is included in $\bm{U}_{k}$ for simplicity of notation.
\end{rem}
By using the presented sampling approach from Section~\ref{sec:prelim_sampling}, we can inner-approximate the set of all states that satisfy the chance constraints~\eqref{eq:statecons} by making sure that every drawn sample trajectory $\bm{X}^{(i)}$ satisfies the hard constraints ${\gls{x}^{(i)}_{l|k}\in\mathbb{X}}~{\forall l \in \mathbb{N}_1^L},\, {i \in \mathbb{N}_1^{N_s}}$, resulting in linear constraints for $\gls{xh}_{k}$ and $\bm{U}_{k}$ due to~\eqref{eq:samplepred}. For this, let us define the confidence level $\beta \in (0,1)$, let $\bm{H}_{\eps}^{(i)}$ be Hankel matrices of independently drawn noise sequences $\gls{eps}_{[1,\,N]}^{\text{d},(i)}$, and let $\gls{eps}_{k}^{(i)}$ be independently drawn noise realizations. We define the set $\mathbb{X}^P_l$ of initial states $\gls{xh}_{k}$ and input sequences $\bm{U}_{k}$ for which all predicted states $\gls{x}_{l|k}$ satisfy the chance constraint~\eqref{eq:statecons} as
\begin{equation} \label{eq:chanceset}
    \mathbb{X}^P_l = \left\{ \left. \vc{\gls{xh}_{k}\\ \bm{U}_{k}}  \,\right|~ \Pr{\bm{G}_x \gls{x}_{l|k} \le \bm{g}_x} \ge p \right\},~~l \in \mathbb{N}_1^L.
\end{equation}
Using~\eqref{eq:samplepred} and Proposition~\ref{prop:sampling}, the sampled set
\begin{equation} \label{eq:sampledset}
    \mathbb{X}^S_l = \left\{ \vc{\gls{xh}_{k}\\ \bm{U}_{k}} \,\left|~ \bm{G}_{x,l}^{(i)} \vc{\gls{xh}_{k}\\ \bm{U}_{k}} \le \bm{g}_{x,l}^{(i)},~i \in \mathbb{N}_1^{N_s}\right. \right\}
\end{equation}
is an inner-approximation of~\eqref{eq:chanceset} with confidence $\beta$ for a given risk parameter $p$ if $N_s \ge \tilde{N}(n+Lm, p,\beta)$, with
\begin{subequations}
    \begin{align}
   \bm{G}_{x,l}^{(i)} &=\bm{G}_x \left[\bm{M}^{(i)}\right]_{[ln+1,\, (l+1)n]},\\
    \bm{g}_{x,l}^{(i)} &= \bm{g}_x - \underline{\bm{G}}_x \left[\bm{M}^{(i)}\right]_{[ln+1,\, (l+1)n]} \vc{-\gls{eps}_k^{(i)}\\ \bm{0}}.
\end{align}
\end{subequations}
We can now intersect the sampled state constraints~\eqref{eq:sampledset} with the input constraints~\eqref{eq:inputcons} for all predicted steps ${l \in \mathbb{N}_1^L}$, and remove redundant constraints from the resulting set, leading to the overall constraint set of the form
\begin{equation} \label{eq:constraints_red}
    \vc{\gls{xh}_{k}\\ \bm{U}_{k}} \in \mathbb{C} = \left\{ \vc{\gls{xh}_{k}\\ \bm{U}_{k}} \left|~ \bm{G} \vc{\gls{xh}_{k}\\ \bm{U}_{k}} \le \bm{g}\right. \right\}.
\end{equation}
Next, we add additional constraints for the first predicted step to later guarantee recursive feasibility, similar to~\cite{lorenzen2016constraint, lorenzen2017stochastic}.

\subsection{First-Step Constraint}
Let us define the set of feasible (noisy) initial states for a prediction horizon of length $L$ as
\begin{equation}
    \mathbb{Z}_L = \left\{ \gls{xh}_{k} \left|~ \begin{array}{c}
         \exists \gls{u}_{0|k}, \dots, \gls{u}_{L-1|k} \in \mathbb{U}:\\
         \vc{\gls{xh}_{k}^{\top}, & \bm{U}^{\top}_{k}}^{\top} \in \mathbb{C}
    \end{array}\right. \right\},
\end{equation}
which can be computed by projection of~\eqref{eq:constraints_red} onto the first $n$ dimensions. Since $\mathbb{Z}_L$ is not necessarily robustly positive invariant regarding the uncertainties $\bm{H}_{\eps}$ and $\gls{eps}_{k}$ due to the probabilistic constraint tightening, we further need to compute a robust control invariant set $\mathbb{Z}_L^{\infty}$ that guarantees
\begin{align*}
     &\forall \gls{eps}_{k}, \gls{eps}_{k+1} \in \glsd{eps} \,~ \exists \gls{u}_{k} \in \mathbb{U}:\\
     &\gls{xh}_{k} \in \mathbb{Z}_L^{\infty} \Longrightarrow \gls{xh}_{k+1} = \bm{A} \gls{xh}_{k} + \bm{B} \gls{u}_{k} - \bm{A} \gls{eps}_{k} + \gls{eps}_{k+1} \in \mathbb{Z}_L^{\infty}.
\end{align*}
According to \cite{blanchini2015set}, the set $\mathbb{Z}_L^{\infty}$ can be determined through $\mathbb{Z}^{\infty}_L = \cap_{q=0}^{\infty} \mathbb{Z}^{q}_L$, with $\mathbb{Z}^{0}_L = \mathbb{Z}_L$ and
\begin{equation}
	\mathbb{Z}^{q+1}_L = \left\{ \gls{xh}_{k} ~\left|~ \begin{array}{l} 
	\forall \gls{eps}_{k},\, \gls{eps}_{k+1} \in \glsd{eps},\, j \in \mathbb{N}_1^{N_c} \,~\exists \gls{u}_{k} \in \mathbb{U}: \\
	\gls{xh}_{k} \in \mathbb{Z}_L \cap \mathbb{Z}^{q}_L, \\
	\bm{A}_j \gls{xh}_{k} + \bm{B}_j \gls{u}_{k} - \bm{A}_j \gls{eps}_{k} + \gls{eps}_{k+1} \in \mathbb{Z}_L^q
	\end{array} \right.
	\right\},	
\end{equation}
where $\bm{A}_j$ and $\bm{B}_j$ are taken from $\mathbb{A}$ (cf. Assumption~\ref{ass:bound}).

In practice, $\mathbb{Z}^{\infty}_L$ is computed by recursively computing $\mathbb{Z}^{q}_L$ until $\mathbb{Z}^{q}_L = \mathbb{Z}^{q+1}_L$ for some $q \in \mathbb{N}$, which implies that ${\mathbb{Z}^{\infty}_L = \mathbb{Z}^{q}_L}$. As $\glsd{eps}$ is polytopic, it is sufficient to only take the vertices of $\glsd{eps}$ into account during computation.

Now, let us express $\mathbb{Z}^{\infty}_L$ as $\mathbb{Z}^{\infty}_L = \left\{\gls{x} \in \mathbb{R}^n ~\left|~ \bm{G}_{\infty} \gls{x} \le \bm{g}_{\infty} \right.\right\}$.
In order to later guarantee recursive feasibility of the control scheme, we construct a constraint set that imposes a robust constraint on the first predicted input $\gls{u}_{0|k}$, i.e.
\begin{equation} \label{eq:constraints_fs}
    \mathbb{C}_R = \left\{ \vc{\gls{xh}_{k}\\ \bm{U}_{k}} ~\left|~ \begin{array}{l}
        \forall \gls{eps}_{k},\, \gls{eps}_{k+1} \in \glsd{eps},\, j \in \mathbb{N}_1^{N_c}: \\
         \bm{G}_{\infty} \bm{A}_j \gls{xh}_{k} + \bm{G}_{\infty} \bm{B}_j \gls{u}_{0|k} \le  \\
        \le\bm{g}_{\infty} - \bm{G}_{\infty}\gls{eps}_{k} + \bm{G}_{\infty}\bm{A}_j \gls{eps}_{k+1}
    \end{array} \right. \right\}.
\end{equation}
Based on the results from all previous sections, we can now state the proposed control algorithm.
\subsection{Control Algorithm}
We define the OCP that is solved at each time instant $k$ for a given state measurement $\gls{xh}_{k}$ as
\begin{subequations} \label{eq:ocp}
	\begin{align}
		\bm{U}^{\ast}_{k} &=\arg\underset{\bm{U}_{k}}{\min} ~
		\vc{\gls{xh}_{k} \\ \bm{U}_{k}}^{\top} \bm{S} \vc{\gls{xh}_{k} \\ \bm{U}_{k}} + \bm{\gamma}^{\top}\vc{\gls{xh}_{k} \\ \bm{U}_{k}} \label{eq:costfunction}\\	
		\text{s.t. } & \vc{\gls{xh}_{k}\\ \bm{U}_{k}} \in \mathbb{C} \cap \mathbb{C}_R.
	\end{align}
\end{subequations}
Note that a solution to \eqref{eq:ocp} can only exist if $\mathbb{C} \cap \mathbb{C}_R$ is non-empty, i.e., the bounds in Assumptions~\ref{ass:noise} and \ref{ass:bound} are suitably tight. The overall proposed control algorithm can be split into an offline and online phase as follows.

\textit{Offline}: Retrieve a persistently exciting (noisy) input-state data trajectory. Determine the weight $\bm{S}$ of the cost function~\eqref{eq:costfunction} via \eqref{eq:weight-S}. Use Proposition~\ref{prop:sampling} to determine the required numbers of samples $N_s$. Determine the sampled constraint set and remove redundant constraints to retrieve~\eqref{eq:constraints_red}, and compute the first-step constraint~\eqref{eq:constraints_fs}\footnote{Algorithms for redundancy removal and set computations as part of a toolbox are provided by~\cite{MPT3}.}.\\
\textit{Online}: In each time step $k \in \mathbb{N}$, measure the current (noisy) state $\gls{xh}_{k}$, solve the OCP~\eqref{eq:ocp}, and apply $\gls{u}_{k} = \gls{u}^{\ast}_{0|k}$, i.e., the first element of the optimal input sequence $\bm{U}_k^{\ast}$.

\subsection{Theoretical Properties}
In the following, we present control theoretic properties of the proposed controller.

\begin{thm}[Recursive Feasibility] \label{th:recfeas}
    Let 
    \begin{equation}
        \mathbb{F}\left(\gls{xh}_{k}\right) = \left\{ \bm{U}_{k} ~\left|~ \vc{\gls{xh}_{k}^{\top}, & \bm{U}^{\top}_{k}}^{\top} \in \mathbb{C} \cap \mathbb{C}_R \right.\right\}
    \end{equation} 
    be the set of all feasible input sequences for a given state measurement~$\gls{xh}_{k}$. If a given optimizer $\bm{U}^{\ast}_{k}$ of~\eqref{eq:ocp} is feasible at time $k$, i.e., $\bm{U}^{\ast}_{k} \in \mathbb{F}\left(\gls{xh}_{k}\right)$, then $\mathbb{F}\left(\gls{xh}_{k+1}\right) \neq \emptyset$ for every realization of the measurement noise $\gls{eps}_{k},\,\gls{eps}_{k+1} \in \glsd{eps}$, with $\gls{xh}_{k+1} = \bm{A} \gls{x}_{k} + \bm{B} \gls{u}_{k} + \gls{eps}_{k+1}$, $\gls{x}_{k} = \gls{xh}_{k} - \gls{eps}_{k}$, and  $\gls{u}_{k} = \gls{u}^{\ast}_{0|k}$.
\end{thm}

\begin{pf}
    By robustness of the first-step constraint~\eqref{eq:constraints_fs}, ${\gls{xh}_{k+1} \in \mathbb{Z}_L^{\infty}}$ follows from $\vc{\gls{xh}_{k}^{\top},& \bm{U}_{k}^{\ast \top}}^{\top} \in \mathbb{C}_R$. Furthermore, ${\mathbb{Z}_L^{\infty} \subset \left\{\gls{x} \,\left|\, \mathbb{F}\left(\gls{x}\right) \neq \emptyset \right.\right\}}$ holds by construction of $\mathbb{Z}_L^{\infty}$, leading to $\mathbb{F}\left(\gls{xh}_{k+1}\right) \neq \emptyset$.
\end{pf}

\begin{prop}[Constraint Satisfaction]
    Under the proposed control law, the closed-loop system satisfies the chance constraint \eqref{eq:statecons} with confidence $\beta$, as well as the input constraints \eqref{eq:inputcons} for all $k \in \mathbb{N}$ if~${\gls{x}_0 \in \mathbb{Z}_L^{\infty} \ominus \glsd{eps}}$.
\end{prop}

\begin{pf}
    For ${\gls{x}_0 \in \mathbb{Z}_L^{\infty} \ominus \glsd{eps}}$, it holds that ${\gls{xh}_0 \in \mathbb{Z}_L^{\infty}}$, and thus, a feasible pair $\vc{\gls{xh}_0^{\top}, & \bm{U}^{\top}_{0}}^{\top} \in \mathbb{C}$ for $k=0$ exists. Closed-loop input constraint satisfaction follows from Theorem~\ref{th:recfeas} and the constraint $\gls{u}_{0|k} \in \mathbb{U}$, which is included in the constraint set $\mathbb{C}$.
    Furthermore, it holds that $\mathbb{C} \subseteq \mathbb{X}_1^S$, and $\mathbb{X}_1^S \subseteq \mathbb{X}_1^P$ with confidence $\beta$ according to Proposition~\ref{prop:sampling}. Thus, the chance constraint $\Pr{\bm{G}_x \gls{x}_{1|k} \le \bm{g}_x} \ge p$ is satisfied with confidence $\beta$ for all feasible $\vc{\gls{xh}_{k}^{\top}, & \bm{U}^{\top}_{k}}^{\top} \in \mathbb{C}$, $k \in \mathbb{N}$, which is sufficient for the chance constraint~\eqref{eq:statecons}.
\end{pf}

\subsection{Discussion}

Similar to \cite{lorenzen2017stochastic}, the proposed method uses an offline-sampling approach to deterministically approximate the probabilistic chance constraints. This allows to remove redundant constraints before applying the control scheme, which reduces the computational load in the online optimization. Furthermore, the offline-sampling approach allows to define a first-step constraint that guarantees robust recursive feasibility. In contrast to \cite{lorenzen2017stochastic}, the proposed method considers measurement noise and does not rely on model knowledge.

The proposed uncertainty-sampling based constraint tightening approach is a novelty for data-driven predictive control, and provides an alternative to conservative robust constraint tightening, e.g., as proposed by~\cite{berberich2020robust}, while robust recursive feasibility is still guaranteed without terminal equality constraints that restrict the region of feasible initial values.

%% file: 5_eval.tex
\section{NUMERICAL EVALUATION} \label{sec:eval}
We now present a simple simulation example that demonstrates the improved performance of the proposed predictive controller with offline uncertainty sampling in comparison to the robust approach by~\cite{berberich2020robust}.

\subsection{Simulation Setup}
We consider the linear system \citep{lorenzen2016constraint}
\begin{subequations}
	\begin{align}
	\gls{x}_{k+1} &= \mat{1 & 0.013 \\ -0.080 & 0.996} \gls{x}_{k} + \mat{4.798 \\ 0.064} \gls{u}_{k}, \\
	\gls{xh}_{k} &= \gls{x}_{k} + \gls{eps}_{k}, \hspace{5mm} k \in \mathbb{N}_0,
	\end{align}
\end{subequations}
where the measurement noise $\gls{eps}_{k} \in \glsd{eps} = \left\{\bm{\eps} \,\left|\, \norminf{\gls{eps}} \le \gls{epsbnd} \right.\right\}$ is the realization of a random variable, distributed according to a zero-mean Gaussian with variance $\gls{epsbnd}/3$ in each dimension,  truncated at the noise bound $\gls{epsbnd}$, for which we consider four scenarios with different noise levels:
$\gls{epsbnd}\in \{0.0001,0.001,0.002,0.01,0.1\}$.
The system matrices are assumed to be unknown and only persistently exciting data in the form of a single (noisy) measured trajectory ${(\gls{ud}_{[1,\,N]},~\gls{xhd}_{[1,\,N]})}$ is available. To retrieve this trajectory, we simulate $N=30$ time steps of the system in open-loop for each noise scenario, with randomly chosen inputs.

The control objective is to track a reference state $\gls{x}_{\text{ref}} = \left(0,\,2.8\right)^{\top}$ while adhering to state constraints ${\norminf{\gls{x}_{k}} \le 2.8}$ and input constraints ${\norminf{\gls{u}_{k}} \le 0.2}$. With risk parameter ${p = 0.8}$ and confidence ${\beta = 0.999}$, we require $N_s = 31\,800$ uncertainty samples via \eqref{eq:samplecomplexity}. After redundancy removal, we are left with a total of $1\,172$ to $2\,380$, depending on the noise scenario. 
For the OCP~\eqref{eq:ocp}, we use a cost function as defined in \eqref{eq:cost_explicit} with prediction horizon ${L = 6}$ and weighting matrices $\bm{R} = 1$, $\bm{P} = \bm{Q} = \diag{1,\,10}$.

In a Monte-Carlo simulation of $1\,000$ runs per noise scenario, the initial state is randomly chosen from $\left\{\gls{x}_{0} \,|\,\right.$ $\left.\norminf{\gls{x}_{0}} \le 0.5\right\}$ and the proposed controller is applied for $30$ time steps, where at each time step the measured state is subject to measurement noise. The simulations are carried out in MATLAB using the \texttt{quadprog} solver. We use the MPT3 toolbox \citep{MPT3} for set computations.

\subsection{Simulation Results}
\begin{figure}
    \centering
    \vspace{-2mm}
    \hspace*{1pt}
    \def\svgwidth{0.48\textwidth}
    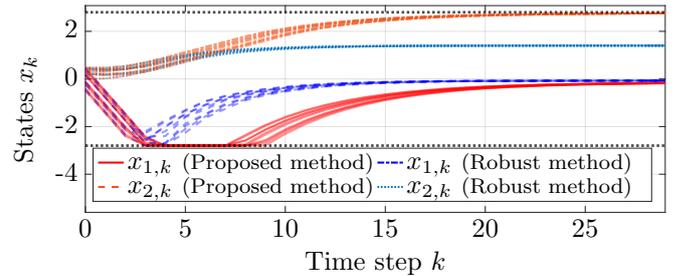
    \vspace{-8mm}
    \caption{Comparison between the proposed method and the robust method by \cite{berberich2020robust}: trajectories of 10 runs with noisy data and online measurements for the noise bound $\gls{epsbnd} = 0.002$. Constraints are shown in dotted black lines.}
    \label{fig:traj}
\end{figure}

In order to appropriately evaluate the level of conservatism, we compare the proposed approach to the purely robust data-driven strategy by \cite{berberich2020robust}. Fig.~\ref{fig:traj} shows 10 exemplary trajectories
of both methods for $\gls{epsbnd} = 0.002$. The robust strategy by \cite{berberich2020robust} results in a large tracking error as the system operates more carefully due to more conservative constraint tightening. In contrast, the probabilistic constraint tightening of the proposed approach allows the system to operate near the constraint boundary for all investigated noise levels, leading to fast and close convergence. No constraint violations occur in any run due to the robust first-step constraint. The mean computation time for solving the OCP~\eqref{eq:ocp} per time step is $\SI{1.18}{\milli\second}$ on an AMD Ryzen 5 Pro 3500U.

\begin{figure}
    \centering
    \def\svgwidth{0.49\textwidth}
    \hspace*{1pt}
    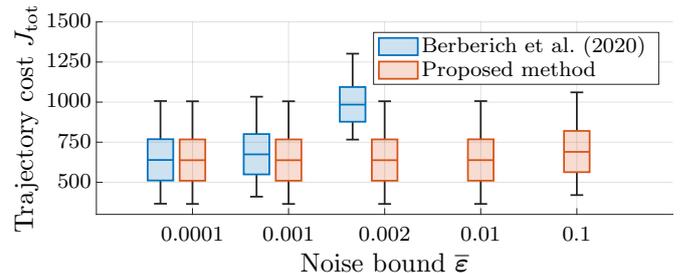
    \vspace{-6mm}
    \caption{Boxplots of the total trajectory cost $J_{\text{tot}}$ for both the proposed method and the robust approach~\citep{berberich2020robust} in different noise scenarios. Note that the robust approach is infeasible for noise bound $\gls{epsbnd} = 0.01$ and $\gls{epsbnd} = 0.1$.}
    \label{fig:costs}
\end{figure}

In Fig.~\ref{fig:costs}, boxplots of the total trajectory costs 
\begin{equation}
    J_{\text{tot}}= \textstyle \sum_{k=0}^{29} \left(\bm{e}_{k+1}^{\top} \bm{Q}  \bm{e}_{k+1} + \gls{u}_{k}^{\top}\bm{R}\gls{u}_{k} \right),
\end{equation}
$\bm{e}_k=\gls{x}_{k}-\gls{x}_{\text{ref}}$, for the different simulation scenarios are shown. In the case of large noise, robust constraint tightening leads to vanishing permissible sets, rendering the OCP infeasible. This is not the case for our proposed approach, where the performance only slowly deteriorates as noise increases: Our accumulated costs for the highest tested noise bound $\gls{epsbnd}=0.1$ are close to those of the robust scheme with noise levels two orders of magnitude lower.


%% file: pics/traj4.pdf_tex
\begingroup%
  \makeatletter%
  \providecommand\color[2][]{%
    \errmessage{(Inkscape) Color is used for the text in Inkscape, but the package 'color.sty' is not loaded}%
    \renewcommand\color[2][]{}%
  }%
  \providecommand\transparent[1]{%
    \errmessage{(Inkscape) Transparency is used (non-zero) for the text in Inkscape, but the package 'transparent.sty' is not loaded}%
    \renewcommand\transparent[1]{}%
  }%
  \providecommand\rotatebox[2]{#2}%
  \newcommand*\fsize{\dimexpr\f@size pt\relax}%
  \newcommand*\lineheight[1]{\fontsize{\fsize}{#1\fsize}\selectfont}%
  \ifx\svgwidth\undefined%
    \setlength{\unitlength}{585.66242368bp}%
    \ifx\svgscale\undefined%
      \relax%
    \else%
      \setlength{\unitlength}{\unitlength * \real{\svgscale}}%
    \fi%
  \else%
    \setlength{\unitlength}{\svgwidth}%
  \fi%
  \global\let\svgwidth\undefined%
  \global\let\svgscale\undefined%
  \makeatother%
  \begin{picture}(1,0.45461344)%
    \lineheight{1}%
    \setlength\tabcolsep{0pt}%
    \put(0,0){\includegraphics[width=\unitlength,page=1]{traj4.pdf}}%
    \put(0.09911354,0.0694086){\makebox(0,0)[t]{\lineheight{1.25}\smash{\begin{tabular}[t]{c}0\end{tabular}}}}%
    \put(0.24969456,0.0694086){\makebox(0,0)[t]{\lineheight{1.25}\smash{\begin{tabular}[t]{c}5\end{tabular}}}}%
    \put(0.40027559,0.0694086){\makebox(0,0)[t]{\lineheight{1.25}\smash{\begin{tabular}[t]{c}10\end{tabular}}}}%
    \put(0.55085661,0.0694086){\makebox(0,0)[t]{\lineheight{1.25}\smash{\begin{tabular}[t]{c}15\end{tabular}}}}%
    \put(0.70143764,0.0694086){\makebox(0,0)[t]{\lineheight{1.25}\smash{\begin{tabular}[t]{c}20\end{tabular}}}}%
    \put(0.85201867,0.0694086){\makebox(0,0)[t]{\lineheight{1.25}\smash{\begin{tabular}[t]{c}25\end{tabular}}}}%
    \put(0.53579892,0.0186968){\makebox(0,0)[t]{\lineheight{1.25}\smash{\begin{tabular}[t]{c}Time step $k$\end{tabular}}}}%
    \put(0,0){\includegraphics[width=\unitlength,page=2]{traj4.pdf}}%
    \put(0.08784425,0.15439096){\makebox(0,0)[rt]{\lineheight{1.25}\smash{\begin{tabular}[t]{r}-4\end{tabular}}}}%
    \put(0.08784425,0.22668295){\makebox(0,0)[rt]{\lineheight{1.25}\smash{\begin{tabular}[t]{r}-2\end{tabular}}}}%
    \put(0.08784425,0.29897493){\makebox(0,0)[rt]{\lineheight{1.25}\smash{\begin{tabular}[t]{r}0\end{tabular}}}}%
    \put(0.08784425,0.37126692){\makebox(0,0)[rt]{\lineheight{1.25}\smash{\begin{tabular}[t]{r}2\end{tabular}}}}%
    \put(0.02,0.26316369){\rotatebox{90}{\makebox(0,0)[t]{\lineheight{1.25}\smash{\begin{tabular}[t]{c}States $x_k$\end{tabular}}}}}%
    \put(0,0){\includegraphics[width=\unitlength,page=3]{traj4.pdf}}%
    \put(0.16250329,0.17253836){\makebox(0,0)[lt]{\lineheight{1.25}\smash{\begin{tabular}[t]{l}$x_{1,k}$ \small (Proposed method)\end{tabular}}}}%
    \put(0,0){\includegraphics[width=\unitlength,page=4]{traj4.pdf}}%
    \put(0.16250329,0.13557432){\makebox(0,0)[lt]{\lineheight{1.25}\smash{\begin{tabular}[t]{l}$x_{2,k}$ \small (Proposed method)\end{tabular}}}}%
    \put(0,0){\includegraphics[width=\unitlength,page=5]{traj4.pdf}}%
    \put(0.58318078,0.17253836){\makebox(0,0)[lt]{\lineheight{1.25}\smash{\begin{tabular}[t]{l}$x_{1,k}$ \small (Robust method)\end{tabular}}}}%
    \put(0,0){\includegraphics[width=\unitlength,page=6]{traj4.pdf}}%
    \put(0.58318078,0.13557432){\makebox(0,0)[lt]{\lineheight{1.25}\smash{\begin{tabular}[t]{l}$x_{2,k}$ \small (Robust method)\end{tabular}}}}%
    \put(0,0){\includegraphics[width=\unitlength,page=7]{traj4.pdf}}%
  \end{picture}%
\endgroup%

%% file: pics/costs2.pdf_tex
\begingroup%
  \makeatletter%
  \providecommand\color[2][]{%
    \errmessage{(Inkscape) Color is used for the text in Inkscape, but the package 'color.sty' is not loaded}%
    \renewcommand\color[2][]{}%
  }%
  \providecommand\transparent[1]{%
    \errmessage{(Inkscape) Transparency is used (non-zero) for the text in Inkscape, but the package 'transparent.sty' is not loaded}%
    \renewcommand\transparent[1]{}%
  }%
  \providecommand\rotatebox[2]{#2}%
  \newcommand*\fsize{\dimexpr\f@size pt\relax}%
  \newcommand*\lineheight[1]{\fontsize{\fsize}{#1\fsize}\selectfont}%
  \ifx\svgwidth\undefined%
    \setlength{\unitlength}{573.99057463bp}%
    \ifx\svgscale\undefined%
      \relax%
    \else%
      \setlength{\unitlength}{\unitlength * \real{\svgscale}}%
    \fi%
  \else%
    \setlength{\unitlength}{\svgwidth}%
  \fi%
  \global\let\svgwidth\undefined%
  \global\let\svgscale\undefined%
  \makeatother%
  \begin{picture}(1,0.39920431)%
    \lineheight{1}%
    \setlength\tabcolsep{0pt}%
    \put(0,0){\includegraphics[width=\unitlength,page=1]{costs2.pdf}}%
    \put(0.25501683,0.045){\makebox(0,0)[t]{\lineheight{1.25}\smash{\begin{tabular}[t]{c}\small 0.0001\end{tabular}}}}%
    \put(0.39678745,0.045){\makebox(0,0)[t]{\lineheight{1.25}\smash{\begin{tabular}[t]{c}\small 0.001\end{tabular}}}}%
    \put(0.53855807,0.045){\makebox(0,0)[t]{\lineheight{1.25}\smash{\begin{tabular}[t]{c}\small 0.002\end{tabular}}}}%
    \put(0.68032869,0.045){\makebox(0,0)[t]{\lineheight{1.25}\smash{\begin{tabular}[t]{c}\small 0.01\end{tabular}}}}%
    \put(0.82209931,0.045){\makebox(0,0)[t]{\lineheight{1.25}\smash{\begin{tabular}[t]{c}\small 0.1\end{tabular}}}}%
    \put(0.53855846,0.0){\makebox(0,0)[t]{\lineheight{1.25}\smash{\begin{tabular}[t]{c}Noise bound $\overline{\gls{eps}}$\end{tabular}}}}%
    \put(0,0){\includegraphics[width=\unitlength,page=2]{costs2.pdf}}%
    \put(0.10592902,0.12417587){\makebox(0,0)[rt]{\lineheight{1.25}\smash{\begin{tabular}[t]{r}\small 500\end{tabular}}}}%
    \put(0.10592902,0.1837914){\makebox(0,0)[rt]{\lineheight{1.25}\smash{\begin{tabular}[t]{r}\small 750\end{tabular}}}}%
    \put(0.10592902,0.24340693){\makebox(0,0)[rt]{\lineheight{1.25}\smash{\begin{tabular}[t]{r}\small 1000\end{tabular}}}}%
    \put(0.10592902,0.30302245){\makebox(0,0)[rt]{\lineheight{1.25}\smash{\begin{tabular}[t]{r}\small 1250\end{tabular}}}}%
    \put(0.10592902,0.36263798){\makebox(0,0)[rt]{\lineheight{1.25}\smash{\begin{tabular}[t]{r}\small 1500\end{tabular}}}}%
    \put(0.02,0.2274007){\rotatebox{90}{\makebox(0,0)[t]{\lineheight{1.25}\smash{\begin{tabular}[t]{c}Trajectory cost $J_{\text{tot}}$\end{tabular}}}}}%
    \put(0,0){\includegraphics[width=\unitlength,page=3]{costs2.pdf}}%
    \put(0.59332661,0.32544037){\makebox(0,0)[lt]{\lineheight{1.25}\smash{\begin{tabular}[t]{l}\small \cite{berberich2020robust}\end{tabular}}}}%
    \put(0,0){\includegraphics[width=\unitlength,page=4]{costs2.pdf}}%
    \put(0.59332661,0.29033901){\makebox(0,0)[lt]{\lineheight{1.25}\smash{\begin{tabular}[t]{l}\small Proposed method\end{tabular}}}}%
    \put(0,0){\includegraphics[width=\unitlength,page=5]{costs2.pdf}}%
  \end{picture}%
\endgroup%

%% file: 6_conclusion.tex
\section{CONCLUSION AND FUTURE WORK} \label{sec:conclusion}
In this paper, we proposed a sampling-based data-driven predictive control scheme for the control of an unknown linear system, where the state measurement is affected by probabilistic noise. A result from behavioral systems theory allows for model-free predictions, only based on past system data. Under the assumption that bounds to the system matrices and measurement noise are known, robust recursive feasibility and closed-loop constraint satisfaction is guaranteed. In the provided simulation example, the proposed controller showed improved control performance regarding cost effectiveness to a purely robust DD-MPC approach, as well as feasibility for noise bounds that were higher by multiple orders of magnitude.

The proposed control scheme provides performance benefits for applications where safe operation near constraint boundaries is required, but the system model is unknown and only noisy data are available. Heavy computations such as the constraint sampling and redundancy removal are done offline, resulting in low online computational cost.

In future work, we will extend the approach to linear systems subject to additive disturbances, and analyze stability of the control scheme. Furthermore, we plan to consider general output definitions.

%% file: appendix.tex
\section{Deriving Outer-bounding set} \label{app:bounds}
Following \cite{de2019formulas}, the system matrices can be retrieved by using noise-free data, i.e.,
\begin{equation} \label{eq:sysid}
    \mat{\bm{A},\,\bm{B}} = \left[\bm{H}_{x}\right]_{[n+1,2n]} \mat{\left[\bm{H}_{x}\right]_{[1,n]}\\ \left[\bm{H}_u\right]_{[1,m]}}^{\dagger}.
\end{equation}
As we only have access to noisy data, \eqref{eq:sysid} cannot be applied. However, using knowledge about the measurement noise (cf. Assumption~\ref{ass:noise}), we can find an outer-bounding set $\mathbb{A}$ to the system matrices. Recall that for any matrix $\bm{D}$ with the elements $d_{ij}$, it holds that $\norm{\bm{D}}_p \ge \max_{i,j}\lvert d_{ij} \rvert$ for any induced $p$-norm with ${1 \le p \le \infty}$.
An upper bound $\hat{\rho} \ge \norm{\mat{\bm{A},\,\bm{B}}}_p$ can be retrieved by solving the optimization 
\begin{subequations} \label{eq:estimate_sysparam}
	\begin{align}
	&\hspace{-2mm}\hat{\rho} = \max\limits_{\overline{\bm{\eps}}_{[1,N]}} \norm{\left[\bm{H}_{\hat{x}} - \bm{H}_{\overline\eps}\right]_{[n+1,2n]} \mat{\left[\bm{H}_{\hat{x}} - \bm{H}_{\overline\eps}\right]_{[1,n]}\\ \left[\bm{H}_u\right]_{[1,m]}}^{\dagger}}_{p} \\		
	&\hspace{5mm}\text{s.t. }  \bm{H}_{\overline\eps} = \bm{H}_{L+1}\left( \overline{\bm{\eps}}_{[1,N]}\right),~~\overline{\gls{eps}}_i \in \glsd{eps}~~ \forall i \in \mathbb{N}_1^N,
	\end{align}
\end{subequations}
using the given data and the known noise bound $\glsd{eps}$. As $\hat{\rho} \ge \lvert \tilde{a}_{ij} \rvert$ holds for any element $\tilde{a}_{ij}$ of $\mat{\bm{A},\,\bm{B}}$, $\pm \hat{\rho}$ can be used to construct matrix vertices of the polytopic set~$\mathbb{A}$.